\documentclass[twocolumn]{aa}
\usepackage{graphicx,epsf}
\begin{document}
\renewcommand{\topfraction}{0.85}
\renewcommand{\bottomfraction}{0.7}
\renewcommand{\textfraction}{0.15}
\renewcommand{\floatpagefraction}{0.90}
   \title{Discovery of VHE Gamma Rays from PKS 2005$-$489}

\author{F. Aharonian\inst{1}
 \and A.G.~Akhperjanian \inst{2}
 \and K.-M.~Aye \inst{3}
 \and A.R.~Bazer-Bachi \inst{4}
 \and M.~Beilicke \inst{5}
 \and W.~Benbow \inst{1}
 \and D.~Berge \inst{1}
 \and P.~Berghaus \inst{6} \thanks{Universit\'e Libre de
 Bruxelles, Facult\'e des Sciences, Belgium}
 \and K.~Bernl\"ohr \inst{1,7}
 \and C.~Boisson \inst{8}
 \and O.~Bolz \inst{1}
 \and I.~Braun \inst{1}
 \and F.~Breitling \inst{7}
 \and A.M.~Brown \inst{3}
 \and J.~Bussons Gordo \inst{9}
 \and P.M.~Chadwick \inst{3}
 \and L.-M.~Chounet \inst{10}
 \and R.~Cornils \inst{5}
 \and L.~Costamante \inst{1,20}
 \and B.~Degrange \inst{10}
 \and A.~Djannati-Ata\"i \inst{6}
 \and L.O'C.~Drury \inst{11}
 \and G.~Dubus \inst{10}
 \and D.~Emmanoulopoulos \inst{12}
 \and P.~Espigat \inst{6}
 \and F.~Feinstein \inst{9}
 \and P.~Fleury \inst{10}
 \and G.~Fontaine \inst{10}
 \and Y.~Fuchs \inst{13}
 \and S.~Funk \inst{1}
 \and Y.A.~Gallant \inst{9}
 \and B.~Giebels \inst{10}
 \and S.~Gillessen \inst{1}
 \and J.F.~Glicenstein \inst{14}
 \and P.~Goret \inst{14}
 \and C.~Hadjichristidis \inst{3}
 \and M.~Hauser \inst{12}
 \and G.~Heinzelmann \inst{5}
 \and G.~Henri \inst{13}
 \and G.~Hermann \inst{1}
 \and J.A.~Hinton \inst{1}
 \and W.~Hofmann \inst{1}
 \and M.~Holleran \inst{15}
 \and D.~Horns \inst{1}
 \and O.C.~de~Jager \inst{15}
 \and B.~Kh\'elifi \inst{1}
 \and Nu.~Komin \inst{7}
 \and A.~Konopelko \inst{1,7}
 \and I.J.~Latham \inst{3}
 \and R.~Le Gallou \inst{3}
 \and A.~Lemi\`ere \inst{6}
 \and M.~Lemoine-Goumard \inst{10}
 \and N.~Leroy \inst{10}
 \and T.~Lohse \inst{7}
 \and O.~Martineau-Huynh \inst{16}
 \and A.~Marcowith \inst{4}
 \and C.~Masterson \inst{1,20}
 \and T.J.L.~McComb \inst{3}
 \and M.~de~Naurois \inst{16}
 \and S.J.~Nolan \inst{3}
 \and A.~Noutsos \inst{3}
 \and K.J.~Orford \inst{3}
 \and J.L.~Osborne \inst{3}
 \and M.~Ouchrif \inst{16,20}
 \and M.~Panter \inst{1}
 \and G.~Pelletier \inst{13}
 \and S.~Pita \inst{6}
 \and G.~P\"uhlhofer \inst{1,12}
 \and M.~Punch \inst{6}
 \and B.C.~Raubenheimer \inst{15}
 \and M.~Raue \inst{5}
 \and J.~Raux \inst{16}
 \and S.M.~Rayner \inst{3}
 \and I.~Redondo \inst{10,20}\thanks{now at Department of Physics and
Astronomy, Univ. of Sheffield, U.K.}
 \and A.~Reimer \inst{17}
 \and O.~Reimer \inst{17}
 \and J.~Ripken \inst{5}
 \and L.~Rob \inst{18}
 \and L.~Rolland \inst{16}
 \and G.~Rowell \inst{1}
 \and V.~Sahakian \inst{2}
 \and L.~Saug\'e \inst{13}
 \and S.~Schlenker \inst{7}
 \and R.~Schlickeiser \inst{17}
 \and C.~Schuster \inst{17}
 \and U.~Schwanke \inst{7}
 \and M.~Siewert \inst{17}
 \and H.~Sol \inst{8}
 \and R.~Steenkamp \inst{19}
 \and C.~Stegmann \inst{7}
 \and J.-P.~Tavernet \inst{16}
 \and R.~Terrier \inst{6}
 \and C.G.~Th\'eoret \inst{6}
 \and M.~Tluczykont \inst{10,20}
 \and G.~Vasileiadis \inst{9}
 \and C.~Venter \inst{15}
 \and P.~Vincent \inst{16}
 \and H.J.~V\"olk \inst{1}
 \and S.J.~Wagner \inst{12}}

   \offprints{Wystan.Benbow@mpi-hd.mpg.de}
 
\institute{
Max-Planck-Institut f\"ur Kernphysik, Heidelberg, Germany;
\and
 Yerevan Physics Institute, Armenia;
\and
University of Durham, Department of Physics, U.K.;
\and
Centre d'Etude Spatiale des Rayonnements, CNRS/UPS, Toulouse, France;
\and
Universit\"at Hamburg, Institut f\"ur Experimentalphysik, Germany;
\and
APC, Paris, France;
\thanks{UMR 7164 (CNRS, Universit\'e Paris VII, CEA, Observatoire de Paris)}
\and
Institut f\"ur Physik, Humboldt-Universit\"at zu Berlin, Germany;
\and
LUTH, UMR 8102 du CNRS, Observatoire de Paris, Section de Meudon, France;
\and
Groupe d'Astroparticules de Montpellier, IN2P3/CNRS, Universit\'e Montpellier II, France;
\and
Laboratoire Leprince-Ringuet, IN2P3/CNRS, Ecole Polytechnique, Palaiseau, France;
\and
Dublin Institute for Advanced Studies, Ireland;
\and
Landessternwarte, K\"onigstuhl, Heidelberg, Germany;
\and
Laboratoire d'Astrophysique de Grenoble, INSU/CNRS, Universit\'e Joseph Fourier, France;
\and
DAPNIA/DSM/CEA, CE Saclay, Gif-sur-Yvette, France;
\and
Unit for Space Physics, North-West University, Potchefstroom, South Africa;
\and
Laboratoire de Physique Nucl\'eaire et de Hautes Energies, IN2P3/CNRS, Universit\'es
Paris VI \& VII, France;
\and
Institut f\"ur Theoretische Physik, Lehrstuhl IV, Ruhr-Universit\"at Bochum, Germany;
\and
Institute of Particle and Nuclear Physics, Charles University, Prague, Czech Republic;
\and
University of Namibia, Windhoek, Namibia;
\and
European Associated Laboratory for Gamma-Ray Astronomy, jointly
supported by CNRS and MPG;
}

   \date{Received 29 March 2005; Accepted 20 April 2005}

   \abstract{
The high-frequency peaked BL Lac PKS 2005$-$489 ($z$=$0.071$) 
was observed in 2003 and 2004 with the H.E.S.S. stereoscopic array of imaging
atmospheric-Cherenkov telescopes in Namibia. A signal was detected
at the 6.7$\sigma$ level in the 2004 observations (24.2 hrs live time), but not in 
the 2003 data set (27.3 hrs live time).  PKS 2005$-$489 is the 
first blazar independently discovered by H.E.S.S. to be an emitter of VHE photons, and only the second
such blazar in the Southern Hemisphere.  The integral flux above 200 GeV observed in 2004 is
(6.9$\pm$$1.0_{stat}$$\pm$$1.4_{syst}$) $\times$ 10$^{-12}$ cm$^{-2}$ s$^{-1}$,
corresponding to $\sim$2.5\% of the flux observed from the Crab Nebula.  
The 99\% upper limit on the flux in 2003,
I($>$200 GeV) $<$ 5.2 $\times$ 10$^{-12}$ cm$^{-2}$ s$^{-1}$, is smaller 
than the flux measured in 2004, suggesting an increased level of activity in 2004.
However, the data show no evidence for significant variability on any time scale less than a year.  
An energy spectrum is measured and is characterized by a very soft power law 
(photon index of $\Gamma=4.0\pm0.4$).

   }

   \maketitle
%

\section{Introduction}

PKS 2005$-$489 was initially discovered as a bright ($>$$0.5$ Jy) radio source at 2.7 GHz 
\cite{discovery_paper} and later identified as a very bright BL Lac object \cite{BL_id}.  It is classified
as a high-frequency peaked BL Lac (HBL) due to its X-ray-to-radio flux ratio \cite{Sambruna_paper}
and because its broadband spectral energy distribution (SED) peaks in the UV.  It has been the target of
several multi-wavelength observation campaigns and is well studied from the radio to
the X-ray regime.  PKS 2005$-$489 was also marginally detected by EGRET at energies greater 
than 100 MeV ~\cite{egret1} and in the GeV regime \cite{egret2}.
It is among the closest ($z$=$0.071$) Southern Hemisphere HBLs \cite{redshift}. Based on its
SED and its proximity, PKS 2005$-$489 is 
viewed as a promising candidate for detection as a
VHE emitter (\cite{luigi_AGN}; \cite{perlman_AGN}; \cite{stecker}).  However,
it has not been previously detected in the VHE regime. The CANGAROO collaboration reported upper 
limits on the flux above 2 TeV in 1993-1994 \cite{cangaroo1}, 
above 1.5 TeV in 1997 \cite{cangaroo2}, above 1.1 TeV in 1999 and above 450 GeV in 2000 \cite{cangaroo3}. 
The University of Durham group has published the most constraining
upper limit (3$\sigma$) on the flux, I($>$400 GeV) $<$ 7.9$\times$ 10$^{-12}$ cm$^{-2}$ s$^{-1}$, based 
on observations made from 1996-1999 with the Mark 6 Telescope \cite{durham}.  As upper limits 
are of limited value for interpreting an SED, the present discovery of VHE gamma-rays from PKS 2005$-$489
yields considerably more insight into the understanding of this object and VHE AGN in general.

\section{H.E.S.S. Detector}
The H.E.S.S. experiment, located in the Khomas Highlands of Namibia
(23$^{\circ}$ 16' 18'' S, 16$^{\circ}$ 30' 1'' E, 1835 m above sea level),
is designed to search for astrophysical $\gamma$-ray emission above $\sim$100 GeV.
The detector consists of a system of four imaging atmospheric-Cherenkov telescopes
(diameter 13 m, focal length 15 m, mirror area 107 m$^{2}$) in a square of 120 m side.  
Each telescope is equipped with a camera that provides a 5$^{\circ}$ field of view (f.o.v.)
and contains 960 individual photomultiplier pixels, subtending
0.16$^{\circ}$ each, with Winston cone light concentrators. A H.E.S.S. camera is triggered
when one of 38 overlapping 64-pixel sectors has a minimum number
of pixels with a signal above a threshold in photoelectrons (PEs) 
coincident in an effective $\sim$1.3 ns trigger window.  Once a camera has triggered, 
a signal is sent out to a central trigger system \cite{cent_trig} which allows for a multiple 
telescope coincidence requirement (presently a minimum of two triggered telescopes). The 
sensitivity of H.E.S.S. 
(5$\sigma$ in 25 hours for a 1\% Crab Nebula flux source at 20$^{\circ}$ zenith angle)
allows for detection of VHE emission from objects such as PKS 2005$-$489 at previously
undetectable flux levels.  More details on H.E.S.S. can be 
found in \cite{HESS1}, \cite{HESS2}, and \cite{HESS3}.
 
\section{Observations}
 
The H.E.S.S. observations of PKS 2005$-$489 in 2003 were made while the system was under construction.
Therefore the data were obtained using different instrument configurations.  Most observations in 2003 were
made using a two-telescope array, with the exception of a small amount of data taken after 
the addition of the third telescope to the array in September 2003. Another 
variation arises because the H.E.S.S. central trigger system
was installed in July 2003.  Before this time two-telescope data were taken with each telescope separately,
and the stereo multiplicity requirement was performed off-line (``Offline Stereo'')
using GPS time stamps.  After the installation of the central trigger system, the 
stereo multiplicity requirement was performed in the hardware (``Online Stereo''). 
As the central trigger system reduced the recording rate considerably,
the camera trigger threshold was lowered (3 pixels $>$ 5.3 PEs vs. 4 pixels $>$ 6.7 PEs).
This increased the rates, while maintaining a reduced system dead time, and ultimately
resulted in a lower energy threshold. All observations made in 2004 use the full four-telescope array.

   \begin{table*}
      \caption{Shown are the configurations of H.E.S.S. with which PKS 2005$-$489 was observed,
        the number of observation runs, the dead time corrected observation time,
        the mean zenith angle of the observations (Z$_{obs}$), the post-cuts energy threshold at Z$_{obs}$,
	the number of on-source and off-source events passing the cuts,
        the normalization for the off-source events, the observed excess from PKS 2005$-$489,
        and the significance of the excess.}
         \label{results}
        \centering
         \begin{tabular}{c c c c c c c c c c c c}
            \hline\hline
            \noalign{\smallskip}
	   Dark & & & & Obs. Time & Z$_{\mathrm{obs}}$ & E$_{\mathrm{th}}$ & & & & & Sig.\\
            Periods & Configuration & N$_{\mathrm{tel}}$ & N$_{\mathrm{runs}}$ & [hrs] & [$^{\circ}$] & [GeV] & On & Off & Norm & Excess & [$\sigma$]\\
            \noalign{\smallskip}
            \hline
            \noalign{\smallskip}
            6/2003 & Offline Stereo & 2 & 22 & 7.9  & 28 & 340 & 194 & 1177 & 0.1567 & 10 & 0.7 \\
            7 \& 8/2003 & Online Stereo & 2 & 43 & 18.6 & 29 & 250 & 628 & 3717 & 0.1585 & 39 & 1.5 \\
            9/2003 & Online Stereo & 3 & 2  & 0.8  & 26 & 240 & 56 & 384 & 0.1574 & $-4$ & $-0.5$ \\
            6, 7, 9 \& 10/2004 & Online Stereo & 4 & 57 & 24.2 & 38 & 300 & 1762 & 9410 & 0.1567 & 288 & 6.7  \\
           \noalign{\smallskip}
	    \hline
           \noalign{\smallskip}
	    Total & & & & & & & & & & & 6.3	\\
            \noalign{\smallskip}
            \hline
       \end{tabular}
   \end{table*}

Table~\ref{results} gives details
of the observations of PKS 2005$-$489 by H.E.S.S. which pass selection criteria which
remove data for which the weather conditions were poor or the hardware was not
functioning properly.  The data were taken in 28 minute runs using {\it Wobble} mode, i.e. 
the source direction is positioned $\pm$0.5$^{\circ}$ relative 
to the center of the f.o.v. of the camera during observations, which
allows for both on-source observations and simultaneous estimation
of the background induced by charged cosmic rays.

\section{Analysis Technique}

The analysis of the data passing the run selection criteria proceeds in the following steps:
First the images are calibrated \cite{calib_paper} and then ``cleaned'' to remove night sky background
noise from the image.
The cleaning is done using a two-stage tail-cut procedure (thresholds: 5 \& 10 PEs).
The moments of the shower image are then parameterized using a Hillas-type analysis \cite{hillas}, 
and the shower geometry is reconstructed using the intersection of image axes, giving 
a typical angular resolution of $\sim$0.1$^{\circ}$ per event
and an average accuracy of $\sim$10 m in the determination of the shower
core location. Only images which exceed a minimum total signal (80 PEs for Online Stereo,
90 PEs for Offline Stereo)
and which pass a distance cut requiring the image center of gravity to be less
than 2$^{\circ}$ from the center of the camera are used in the reconstruction.
The size cut (minimum total signal) ensures that the images are well reconstructed and the
distance cut eliminates effects from truncation of images by the
edge of the camera.

After the event reconstruction, the much more numerous cosmic-ray background events
are rejected using cuts on mean reduced scaled width (MRSW)
and length (MRSL) parameters \cite{pks2155_paper}.
A cut on $\theta^{2}$, the square of the angular difference between 
the reconstructed shower position and
the source position, is also applied.
All the cuts (shown in Table~\ref{thecuts}) are optimized {\it a priori} (simultaneously) 
using Monte Carlo gamma-ray simulations and off-source data
to yield the maximum expected significance per hour of observation for a weak source.  
However, the significance expected (and observed) is not strongly dependent on the
exact values of the cuts.

   \begin{table}
      \caption{The selection cuts applied to the data.  The same cuts are used, 
regardless of the number of telescopes, for all Online Stereo configurations.}
         \label{thecuts}
         \centering
         \begin{tabular}{c   c   c   c   c   c }
            \hline\hline
            \noalign{\smallskip}
            Stereo & MRSL & MRSL & MRSW & MRSW & $\theta^2$ \\
            Config. & min & max & min & max \\
            & [$\sigma$] & [$\sigma$] & [$\sigma$] & [$\sigma$] & [deg$^{2}$]\\
            \noalign{\smallskip}
            \hline
            \noalign{\smallskip}
            Offline & 1.3 & $-2.2$ & 1.1 & $-2.5$ & 0.02   \\
            Online & 2.0 & $-2.0$   & 0.9 & $-2.0$ & 0.0125 \\
            \noalign{\smallskip}
            \hline
         \end{tabular}
   \end{table}
   
The background is estimated using all events passing 
cuts in a ring (central radius 0.5$^{\circ}$) around the source location. The ring width is
chosen such that the area of the ring is approximately seven times the area of the on-source region
(i.e. the area, centered on the source location, falling within the $\theta^2$ cut). 
The relative statistical error on the background measurement is reduced by the use of a larger background
region.  A small correction accounting for the radial acceptance of the camera is applied
to the normalization of the off-source event total.  The significance of any excess 
is calculated following the method of Equation (17) in \cite{lima}.

\section{Results}

\begin{figure}
   \centering
      \includegraphics[width=8.7cm]{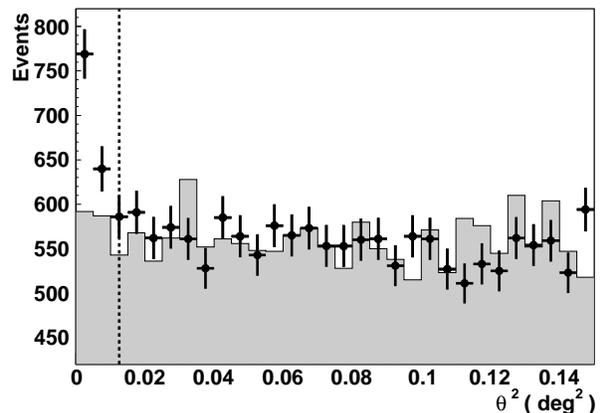} \\ [-0.3cm]
      \caption{The distribution of $\theta^2$ for on-source events (points) and
	normalized off-source events (shaded) from observations of PKS 2005$-$489 in 2004. 
	The dashed line represents the cut on $\theta^2$ applied to the data.}
         \label{thtsq_plot}
   \end{figure}

Table~\ref{results} shows the results of the H.E.S.S. observations for each of the detector
configurations.  A significant excess of events in the direction of PKS 2005$-$489 
is detected in 2004 ($6.7$$\sigma)$, but not in 2003 ($1.4$$\sigma)$.
The total significance of the excess for all observations is 6.3$\sigma$. 
Figure~\ref{thtsq_plot} shows the on-source and normalized off-source
distributions of $\theta^{2}$ for all observations in 2004. The background is
flat in $\theta^{2}$ as expected, and there is a clear excess at small values of
$\theta^{2}$ corresponding to the observed signal. 
A two-dimensional fit of the excess observed finds the 
shape to be characteristic of a point source, located (J2000) at
($\alpha$=$20^{\mathrm h}9^{\mathrm m}29.3^{\mathrm s}$$\pm$$2.7^{\mathrm s}_{stat}$$\pm$$1.3^{\mathrm s}_{syst}$, 
$\delta$=$-48^{\circ}49'19''$$\pm$$36''_{stat}$$\pm$$20''_{syst}$).  The excess, named HESS J2009$-$488,
is consistent with the position of the blazar
($\alpha$=$20^{\mathrm h}9^{\mathrm m}25.4^{\mathrm s}$, $\delta$=$-48^{\circ}49'53.7''$) as expected, and 
is therefore assumed to be associated with PKS 2005$-$489.

\begin{figure}
   \centering
      \includegraphics[width=8.7cm]{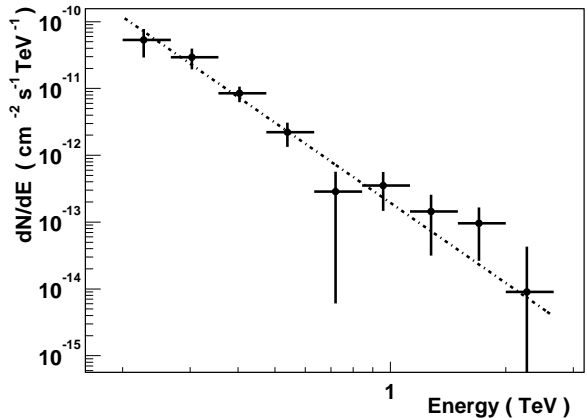} \\ [-0.3cm]
      \caption{The energy spectrum of HESS J2009$-$488.  The dashed line
	represents the best $\chi^2$ fit of a power law.}
         \label{spectrum_plot}
   \end{figure}

As the energy resolution of H.E.S.S. is $\sim$15\% per event, accurate energy spectra can be
measured.  The energy spectrum for the 2004 data set is shown in Figure~\ref{spectrum_plot}.
The best $\chi^2$ fit of a power law ($dN/dE \sim E^{-\Gamma}$) to these data
yields a photon index $\Gamma$=4.0$\pm$$0.4_{stat}$, and a $\chi^2$ of 5.6 for 7 degrees of freedom. 
The systematic error on the photon index is small compared to the statistical error. 
It should be noted that each of the five highest energy points, 
E $>$ 0.64 TeV, in Figure~\ref{spectrum_plot} have statistical significance less than 2$\sigma$.  
However, removing these points from the fit does not change $\Gamma$ significantly. 
Consistent results are also found using alternative background estimation
techniques and/or independent analysis chains.  No evidence is found for significant features, 
such as a cutoff or break, in the energy spectrum.  

Assuming the determined photon index of $\Gamma$=4.0, 
the integral flux above 200 GeV measured in 2004 is
I($>$200 GeV) = (6.9$\pm$$1.0_{stat}$$\pm$$1.4_{syst}$) $\times$ 10$^{-12}$ cm$^{-2}$ s$^{-1}$.  
This corresponds to $\sim$2.5\% of I($>$200 GeV) determined by H.E.S.S. from the Crab Nebula.
The 99\% confidence limit \cite{UL_tech} on I($>$200 GeV) in 2003 is
5.2$\times$ 10$^{-12}$ cm$^{-2}$ s$^{-1}$.  This is slightly less than the
flux in 2004 suggesting that PKS 2005$-$489 was in a higher state during the 2004
observations. The flux upper limit in 2003 and the flux in 2004 are well below
all previously published upper limits for this object.  No evidence for variability in 2003 or 
2004 is found as fits to the integral flux versus time are consistent with being constant.
This is the case whether the data are binned by dark period (months) within 
each year ($\chi^2$ probability, P($\chi^2$), of 0.57 in 2003, and 0.98 in 2004), 
by nights within each dark period (P($\chi^2$)$>$ 0.2 and 0.6 for each of the four periods in 2003 and
2004, respectively) or runs ($\sim$30 min) within 
individual nights (P($\chi^2$)$>$ 0.05 and 0.1 for all nights in 2003 (mean P($\chi^2$) = 0.45)
and 2004 (mean P($\chi^2$) = 0.53), respectively).  
However, given the low statistics overall, the lack of observed short time scale variability 
in 2003 and 2004 is not surprising.

\section{Conclusions}

PKS 2005$-$489 has been detected by H.E.S.S. at energies greater than 200 GeV
in 2004. It is the first AGN independently discovered by H.E.S.S. as an emitter 
of VHE photons and only the second such AGN known in the Southern Hemisphere.
The measured VHE flux is quite low ($\sim$2.5\% of the Crab Nebula flux)
and no evidence supporting variability of the VHE flux on time scales of less than
a year is found.  However, the upper limit resulting from the lack of a detection
in 2003 suggests that the flux from PKS 2005$-$489 in 2004 was higher than
the previous year.  This inference is supported by the behavior of this blazar in the
X-ray regime.  Quick-look results provided by the ASM/RXTE team show the average count
rate from PKS 2005$-$489 was a factor of $\sim$3 higher in 2004 (0.116$\pm$0.025 s$^{-1}$) 
than in 2003 (0.039$\pm$0.026 s$^{-1}$).  Interestingly, the average ASM count rate in 1998 
(0.39$\pm$0.02 s$^{-1}$) is considerably higher than that in 2004, 
suggesting that PKS 2005$-$489 was in a low state during the
H.E.S.S. observations.  Should the VHE flux increase comparably 
to the historical (1998 vs 2004) X-ray count rate, a significant signal 
will quickly accumulate ($\sim$1 hour) in H.E.S.S. observations allowing for more 
detailed studies of the VHE behavior to be performed.  

The VHE spectrum of PKS 2005$-$489 is the softest ($\Gamma$=4.0) ever measured from a BL Lac.
Given the proximity ($z$=$0.071$) of PKS 2005$-$489, the softness is 
unlikely to be largely due to absorption of VHE photons on
the extragalactic background light.  Assuming the softness to be intrinsic to the blazar,
inverse-Compton models of the VHE emission predict that the X-ray spectrum should also be steep.
A multi-wavelength observation campaign (including X-ray energies) was performed in October 2004,
results of which will address this issue among others, but is beyond the scope of this letter.

Given its low flux and soft spectrum, PKS 2005$-$489 was not 
detectable by previous generations of VHE instruments. The currently 
unprecedented ability of H.E.S.S. to detect faint soft-spectrum sources
of VHE gamma-rays, such as AGN, should 
significantly improve the overall understanding of blazars and their physics.

\begin{acknowledgements}
The support of the Namibian authorities and of the University of Namibia
in facilitating the construction and operation of H.E.S.S. is gratefully
acknowledged, as is the support by the German Ministry for Education and
Research (BMBF), the Max-Planck-Society, the French Ministry for Research,
the CNRS-IN2P3 and the Astroparticle Interdisciplinary Programme of the
CNRS, the U.K. Particle Physics and Astronomy Research Council (PPARC),
the IPNP of the Charles University, the South African Department of
Science and Technology and National Research Foundation, and by the
University of Namibia. We appreciate the excellent work of the technical
support staff in Berlin, Durham, Hamburg, Heidelberg, Palaiseau, Paris,
Saclay, and in Namibia in the construction and operation of the
equipment.
\end{acknowledgements}

\end{document}